\UseRawInputEncoding
\documentclass[aps,prd,reprint,amsmath,amssymb,showpacs,superscriptaddress,nofootinbib]{revtex4-2}
\usepackage{epsfig,slashed,xcolor,bm,natbib,comment,enumitem,ulem,cancel}
\usepackage[cal=boondox]{mathalfa}
\def\matri#1{{\mbox{\sf  #1}}}

\def\nd#1#2{{\frac{d #1}{d #2}}}
\def\hnd#1#2#3{{\frac{d^{#3} #1}{d #2 ^{#3}}}}

\def\tfrac#1#2{{\textstyle\frac{#1}{#2}}}

\newcommand{\vpsi}{\varphi}

\newcommand{\comma}{, }
\newcommand{\be}{\begin{equation}}
\newcommand{\ee}{\end{equation}}
\newcommand{\bea}{\begin{eqnarray}}
\newcommand{\eea}{\end{eqnarray}}
\newcommand{\myDel}[1]{{\color{red}\ifmmode\cancel{#1}\else\sout{#1}\fi}}

\newcommand{\chssk}[2]{\genfrac{\{}{\}}{0pt}{}{#1}{#2}}
\begin{document}
\title{Conformally-rescaled Schwarzschild metrics do not
  predict flat galaxy rotation curves} \author{M.P.~\surname{Hobson}}
\email{mph@mrao.cam.ac.uk} \affiliation{Astrophysics Group, Cavendish
  Laboratory, JJ Thomson Avenue, Cambridge CB3 0HE\comma UK}
\author{A.N.~\surname{Lasenby}} \email{a.n.lasenby@mrao.cam.ac.uk}
\affiliation{Astrophysics Group, Cavendish Laboratory, JJ Thomson
  Avenue, Cambridge CB3 0HE\comma UK} \affiliation{Kavli Institute for
  Cosmology, Madingley Road, Cambridge CB3 0HA, UK}
%\date{Received 16 January 2020; accepted ???; published online ???}

\begin{abstract}
For conformally invariant gravity theories defined on Riemannian
spacetime and having the Schwarzschild--de-Sitter (SdS) metric as a
solution in the Einstein gauge, we consider whether one may
conformally rescale this solution to obtain flat rotation curves, such
as those observed in galaxies, without the need for dark
matter. Contrary to recent claims in the literature, we show that if
one works in terms of quantities that can be physically measured, then
in any conformal frame the trajectories followed by `ordinary' matter
particles are merely the timelike geodesics of the SdS metric, as one
might expect. This resolves the apparent frame dependence of physical
predictions and unambiguously yields rotation curves with no flat
region. We also show that attempts to model rising rotation curves by
fitting the coefficient of the quadratic term in the SdS metric
individually for each galaxy are precluded, since this coefficient is
most naturally interpreted as proportional to a global cosmological
constant.  We further extend our analysis beyond static,
spherically-symmetric systems to show that the invariance of particle
dynamics to the choice of conformal frame holds for arbitrary metrics,
again as expected. Moreover, we show that this conclusion remains
valid for conformally invariant gravity theories defined on more
general Weyl--Cartan spacetimes, which include Weyl, Riemann--Cartan
and Riemannian spacetimes as special cases.
\end{abstract}

%\pacs{04.20.−q, 04.50.Kd, 04.80.Cc, 11.30.−j, 95.10.Ce, 95.30.Sf, 98.35.Df}
%04.20.−q Classical general relativity
%04.50.Kd Modified theories of gravity, Gauge field theories, 
%04.80.Cc Experimental tests of gravitational theories
%11.30.−j Symmetry and conservation laws
%95.10.Ce Celestial mechanics
%95.30.Sf Relativity and gravitation
%98.35.Df Kinematics, dynamics, and rotation
\maketitle
%%%%%%%%%%%%%%%%%%%%%%%%%%%%%%%%%%%%%%%%%%%%%%%%%%%%%%%%%%%%%%%%%%%%%%%%%%

The modelling of galaxy rotation curves in general relativity (GR)
typically requires the inclusion of a dark matter halo in order to
reproduce observations \cite{Lelli2016,Lelli2017,Li2020}. 
The family of rotation curves fitted to
observations is, in fact, quite varied \cite{Salucci2019}, but
particular focus has historically been placed on modelling the
approximately flat rotation curves observed in the outskirts of large
spiral galaxies and, to a lesser extent, the rising rotation curves
observed in smaller dwarf galaxies
\cite{Rubin1970,Rubin1978,Rubin1980,Bosma1981,vanAlbada1985}.
The absence
of any direct experimental evidence for dark matter \cite{Feng2010},
however, has led to the consideration of various modified gravity
theories, which may not require a dark matter component to explain the
astrophysical data.  

In its simplest form, the modelling of rotation curves in the
outskirts of galaxies may be performed merely by considering the
motion of stars in the region exterior to a
spherically-symmetric representation of the galactic matter
distribution.
%A&E extension 8758
In any metric-based gravity theory, this simplified approach therefore
considers the motion of massive test particles in a spacetime with line
element that can be written in the form\footnote{We adopt the following sign conventions:
  $(+,-,-,-)$ metric signature, 
%${R^\rho}_{\sigma\mu\nu} = 
%2(\partial_{[\mu}{\Gamma^\rho}_{|\sigma|\nu]}
%+{\Gamma^\rho}_{\lambda[\mu}{\Gamma^\lambda}_{|\sigma|\nu]})$, 
${R^\rho}_{\sigma\mu\nu} = 
2(\partial_{[\mu}\chssk{\rho}{\nu]\sigma}
+\chssk{\rho}{\lambda[\mu}\chssk{\lambda}{\nu]\sigma})$, 
where 
the metric (Christoffel) connection 
%{\Gamma^\rho}_{\lambda\mu}
$\chssk{\rho}{\lambda\mu}=\tfrac{1}{2}g^{\rho\sigma}(\partial_\lambda
g_{\mu\sigma}+\partial_\mu g_{\lambda\sigma}-\partial_\sigma
g_{\lambda\mu})$, and
${R^\rho}_{\mu} = {R^{\rho\sigma}}_{\mu\sigma}$. We also
employ natural units $c=\hbar=1$ throughout,
unless otherwise stated.}
\be
ds^2 = A(r)\,dt^2 - \frac{dr^2}{B(r)} - r^2(d\theta^2 + \sin^2\theta\,d\phi^2),
\label{eqn:genform}
\ee
for given functions $A(r)$ and $B(r)$. 

In GR with a cosmological constant $\Lambda$, the relevant
line-element is Schwarzschild--de-Sitter (SdS), for which
\be
A(r) = B(r) = 1 - \frac{2GM}{r} - kr^2,
\label{eqn:SdSsoln}
\ee
where $M$ is the galactic mass interior to the test particle orbit and
$k=\frac{1}{3}\Lambda$, which is therefore a global constant unrelated
to the galaxy under consideration. Particle rest masses in GR can be
defined kinematically, so that massive (test) particles merely follow
timelike geodesics of the SdS metric. In this case, for a circular
orbit of coordinate radius $r$ (in the equatorial plane $\theta =
\pi/2$), the velocity $v$ of the test particle (as measured by a
stationary observer at that radius) satisfies
\be
v^2  
= \frac{r}{2B}\nd{B}{r}
%\nonumber \\
 =  \frac{GMr^{-1}-k r^{2}}{
1-2GMr^{-1}-kr^{2}}.
\label{eqn:SdSrotcurve}
\ee
In the weak-field limit appropriate for considering a galaxy rotation
curve one has $B \approx 1$, so the two terms in the numerator
determine its shape \cite{Horne2016,Hobson:cgrotpaper}. The first term
recovers the standard Keplerian rotation curve $v^2 = GM/r$ and the
second term contributes $v^2 = -kr^2 = -\frac{1}{3}\Lambda r^2$, so
that for a typical galaxy with $M \sim 10^{11}$~M$_\odot$ and assuming
$\Lambda \sim 10^{-52}$ m$^{-2}$, which is consistent with
cosmological observations, one obtains a rotation curve that falls for
all values of $r$ until bound circular orbits are eliminated beyond
the watershed radius $r = (3GM/\Lambda c^2)^{1/3} \sim 0.5$~Mpc
\cite{Nandra2012}.  Thus, the rotation curve has no flat region.

If one instead considers a conformally invariant gravity theory
defined on Riemannian spacetime, then particle rest masses cannot be
fundamental, but must arise dynamically \cite{Dirac1973}. This may be
achieved by introducing a Dirac field with Weyl weight $w = -3/2$ to
represent `ordinary' matter, which is (Yukawa) coupled to a
compensator scalar field $\vpsi$ with $w=-1$ by making the replacement
$m\bar{\psi}\psi \to \mathcal{m}\vpsi\bar{\psi}\psi$ in the Dirac
action, where $\mathcal{m}$ is a dimensionless parameter but
$\mathcal{m}\vpsi$ has the dimensions of mass in natural units.
Indeed, the full matter action in such theories is usually taken to
have the form \cite{Dirac1973,Mannheim1993a,Mannheim2006,Edery2006}
\bea
S_{\rm M}  &=&  \int d^4x\,\sqrt{-g}\,
[\tfrac{1}{2}i\bar{\psi}\gamma^\rho\,\overleftrightarrow{D}_{\!\!\rho}\psi
  - \mathcal{m}\vpsi\bar{\psi}\psi \nonumber\\
&&
\qquad + \tfrac{1}{2}(\partial_\rho\vpsi) \,(\partial^\rho\vpsi)
- \lambda\vpsi^4 +\tfrac{1}{12}\vpsi^2R],
\label{eqn:sipgt0lm}
\eea
where $\lambda$ is another dimensionless parameter and the numerical
factors ensure that $S_{\rm M}$ varies only by a surface term under a
conformal transformation $g_{\mu\nu} \to \Omega^2(x) g_{\mu\nu}$, for
which $\vpsi \to \Omega^{-1}(x)\vpsi$ and $\psi \to
\Omega^{-3/2}(x)\psi$, where $\Omega(x)$ is any smooth positive
function. In the kinetic term for the Dirac field in
(\ref{eqn:sipgt0lm}), we define
$\bar{\psi}\gamma^\rho\overleftrightarrow{D}_{\!\!\rho}\psi \equiv
\bar{\psi}\gamma^\rho D_{\rho}\psi-(D_\rho\bar{\psi})\gamma^\rho\psi$,
where the spinor covariant derivative has the form ${D_\mu}\psi =
(\partial_\mu + \Gamma_\mu)\psi$, the fermion spin connection
$\Gamma_\mu = \tfrac{1}{8}([\gamma^\lambda,\partial_\mu\gamma_\lambda]
- \chssk{\lambda}{\nu\mu}[\gamma^\nu,\gamma_\lambda])$ and the
position-dependent quantities $\gamma_\mu = {e^a}_\mu\gamma_a$ are
related to the standard Dirac matrices ${\gamma}_a$ using the tetrad
components ${e^a}_\mu$. The total action is then $S_{\rm T} = S_{\rm
  G} + S_{\rm M}$, where the free gravitational contribution $S_G$
depends only on the metric. In Riemannian spacetime, the unique
conformally invariant quadratic action is $S_{\rm G} = \alpha \int
d^4x\,\sqrt{-g}\,C_{\rho\sigma\mu\nu}\,C^{\rho\sigma\mu\nu}$, where
$\alpha$ is a further dimensionless parameter and
$C_{\rho\sigma\mu\nu}$ is the Weyl tensor. The resulting action $S_T$
then describes so-called conformal gravity (also known as Weyl or
Weyl-squared gravity) \cite{Bach1921,Mannheim2006} coupled to Dirac
matter and a compensator scalar field; the special case for which
$\alpha=0$ and $\psi = 0$ is often described as Einstein conformal
gravity \cite{Dirac1973,Englert1976}. Alternative local
\cite{Modesto2016a,Modesto2016b} or non-local
\cite{Krasnikov1987,Modesto2012,Modesto2014} higher-derivative
conformally invariant gravitational actions $S_G$ in Riemannian
spacetime have also been proposed.

In any case, the introduction of the compensator scalar field $\vpsi$
is usually considered important for providing a means for
spontaneously breaking the scale symmetry. Most commonly one uses
local scale invariance to set the scalar field to a constant value
$\vpsi = \vpsi_0$, which is often termed the Einstein gauge. This is
usually interpreted as choosing some definite scale in the theory,
thereby breaking scale-invariance.  As we show in
\cite{Lasenby:eWGTpaper}, however, this interpretation is
questionable, since in such scale-invariant gravity theories the
equations of motion in the Einstein gauge are identical in form to
those obtained when working in scale-invariant variables, which
involves no breaking of the scale symmetry. This suggests that one
should introduce further scalar fields, in addition to the compensator
field $\vpsi$, to enable a true physical breaking of the scale
symmetry. The primary role of the compensator field $\vpsi$ arises
instead from its necessary inclusion into the calculation of physical
quantities, which renders them invariant under local scale
transformations.

In this note, we consider {\it any} conformally invariant gravity
theory in Riemannian spacetime with the matter action
(\ref{eqn:sipgt0lm}) that has the SdS metric (\ref{eqn:SdSsoln}) as a
solution in the Einstein gauge (this includes conformal
gravity). As shown in \cite{Hobson:cgrotpaper}, in this gauge the
scalar field energy-momentum tensor derived from the matter action
(\ref{eqn:sipgt0lm}) vanishes only if $R_{\mu\nu} -
\tfrac{1}{2}g_{\mu\nu}R + 6\lambda\vpsi_0^2g_{\mu\nu} = 0$, so that
the only vacuum metric allowed (assuming that $\psi=0$, apart from
matter test particles) has the SdS form (\ref{eqn:SdSsoln}) with $k =
-2\lambda\vpsi_0^2$.  Thus, in conformally-invariant gravity theories,
unlike GR, the constant $k$ in (\ref{eqn:SdSsoln}) may be system
dependent, if one assumes that $\vpsi_0$ may be so. Hence, there
exists the possibility of attempting to model some (typically rising)
rotation curves by using the expression (\ref{eqn:SdSrotcurve}) to fit
for (negative values of) $k$ separately for each galaxy, as in
\cite{Mirabotalebi2008}. Such an assumption seems questionable when
viewed in the Einstein gauge, however, where $\vpsi_0$ is more
naturally interpreted as a system-independent quantity that leads to a
`global' cosmological constant $\Lambda = -6\lambda\vpsi_0^2$. In this
case, one may therefore no longer fit for $k$ separately for each
galaxy, or at all if one considers $\Lambda$ to be fixed by
cosmological observations. It is also worth noting that, to obtain a
positive cosmological constant $\Lambda$, one must have $\lambda <0$,
which requires a negative scalar field vacuum energy
$\lambda\vpsi^4_0$, at least with the usual sign conventions in the
matter action (\ref{eqn:sipgt0lm}).

Turning to the dynamics of matter test particles, in the Einstein
gauge the rest mass $m=\mathcal{m}\vpsi_0$ of Dirac particles is
independent of spacetime position and so they follow timelike
geodesics of the SdS metric, hence yielding rotation curves with no
flat region.  It has been suggested in \cite{Li2019,Modesto2021},
however, that in such theories one may perform a conformal
transformation of this solution to a frame in which the orbital
velocity of a massive particle in a circular orbit is asymptotically
constant, thereby yielding a flat rotation curve in the outskirts of
galaxies.  Nonetheless, since such theories are (by construction)
conformally invariant, such a transformation should not change the
observable predictions, unless the conformal symmetry is broken in
some way, either dynamically or by imposing boundary
conditions. Merely rescaling the SdS solution to an alternative
conformal frame (or scale gauge) in which the compensator scalar field
$\vpsi$ no longer takes a constant value should preserve the
predictions for physically measurable quantities, such as a rotation
curve. We now demonstrate that this is indeed the case.

As we discuss in \cite{Lasenby:eWGTpaper,Hobson:cgrotpaper}, one may
construct an appropriate action for a spin-$\tfrac{1}{2}$ point
particle and then transition to the full classical approximation in
which the particle spin is neglected. In the presence of the above
Yukawa coupling of the Dirac field to the scalar compensator field
$\vpsi$, this action is equivalent \cite{Hobson:scepaper2} to the standard
action for a massive particle conformally coupled to the scalar field
$\vpsi$, namely
\be
S_{\rm p} = -\mathcal{m} \int d\xi\,\vpsi \sqrt{g_{\mu\nu}\nd{x^\mu}{\xi}
\nd{x^\nu}{\xi}},
\label{ppaction3}
\ee
where $\xi$ is a parameterisation for which the length (squared) $u^2 \equiv
u^\mu u_\mu$ of the tangent vector $u^\mu = dx^\mu/d\xi$ remains
equal to unity along the worldline.

Assuming a static, spherically-symmetric system with $\vpsi=\vpsi(r)$
and a line-element of the form (\ref{eqn:genform}), one finds
that for a massive particle worldline in the
equatorial plane $\theta = \pi/2$, the $t$- and $\phi$-equations of
motion are 
\be
A\Omega^{-1}\nd{t}{\xi} = \mathcal{k},\qquad
r^2\Omega^{-1}\nd{\phi}{\xi} = \mathcal{h},
\label{eqn:parteoms1}
\ee
where $\mathcal{k}$ and $\mathcal{h}$ are constants, and
we may replace the $r$-equation of motion with the much simpler first
integral $u^\mu u_\mu= 1$, which reads
\be
A\left(\nd{t}{\xi}\right)^2 -B^{-1}\left(\nd{r}{\xi}\right)^2 
- r^2\left(\nd{\phi}{\xi}\right)^2 = 1.
\label{eqn:parteoms2}
\ee
Here $\vpsi(r) = \Omega^{-1}(r)\vpsi_0$ 
and the constants
$\mathcal{k}$ and $\mathcal{h}$ are
defined such that one recovers the familiar timelike geodesic
equations in GR for an affine parameter $\xi$ if $\vpsi(r) =
\vpsi_0$ and so $\Omega =1$.

As discussed in
\cite{Hobson:scepaper1,Hobson:cgrotpaper,Hobson:scepaper2}, however,
the parameter $\xi$ cannot be interpreted as the particle proper time,
since it has Weyl weight $w(\xi) = 1$ and so it is not invariant under
conformal transformations. Rather, the proper time interval is instead
given by $d\tau \propto \vpsi\,d\xi$, which is correctly invariant
under conformal transformations. Indeed, one sees from
(\ref{ppaction3}) that the particle dynamics obeys a geodesic
principle, but one where $\vpsi$ must be included in the definition of
the path length to be extremised. Without loss of generality, one may
choose the constant of proportionality such that $d\tau =
(\vpsi/\vpsi_0)\,d\xi = \Omega^{-1}\,d\xi$, so $d\tau$ and $d\xi$
coincide if $\vpsi(r) = \vpsi_0$. When expressed in terms of the
proper time $\tau$ of the particle, and denoting $d/d\tau$ by an
overdot, the equations of motion
(\ref{eqn:parteoms1}--\ref{eqn:parteoms2}) become
\be
A\Omega^{-2}\dot{t} = \mathcal{k},\quad
r^2\Omega^{-2}\dot{\phi} = \mathcal{h},\quad
A\dot{t}^2 -B^{-1}\dot{r}^2 
- r^2\dot{\phi}^2 = \Omega^2.
\label{eqn:parteoms3}
\ee
%

\begin{comment}
On substituting the first two equations
into the third, one straightforwardly obtains the `energy' equation
for massive particle trajectories,
%
\be
\dot{r}^2AB^{-1}\Omega^{-4} + \left(\Omega^{-2}+\frac{\mathcal{h}^2}{r^2}\right)A
= \mathcal{k}^2.
\label{eqn:energy}
\ee
%
Then substituting $\dot{r} = \dot{\phi}\,dr/d\phi =
(\mathcal{h}\Omega^2/r^2)\,dr/d\phi$, defining the reciprocal radial variable $u
\equiv 1/r$ and differentiating with respect to $\phi$, one obtains
the `shape' (or `orbit') equation for massive particle trajectories,
%
\bea 
2\hat{A}\hat{B}^{-1}\hnd{u}{\phi}{2} &+&
\nd{}{u}(\hat{A}\hat{B}^{-1})\left(\nd{u}{\phi}\right)^2\nonumber \\
&+& \nd{}{u}\left[\left(u^2 +
  \frac{1}{\mathcal{h}^2\hat{\Omega}^2}\right)\hat{A}\right] = 0,
\label{eqn:shape}
\eea
%
where we have defined the functions $\hat{A}(u) \equiv A(1/u)$,
$\hat{B}(u) \equiv B(1/u)$ and $\hat{\Omega}(u) \equiv
\Omega(1/u)$. As expected, if $\Omega=1$ the equations
(\ref{eqn:parteoms3}--\ref{eqn:shape}) reduce to the familiar
equations for a timelike geodesic in the equatorial plane
$\theta=\pi/2$ of a static, spherically-symmetric system with
line-element of the form (\ref{eqn:MKform}) \cite{GRbook}.
\end{comment}

If there exists a conformal frame in which a solution for a
static, spherically-symmetric system is given by the metric
(\ref{eqn:genform}) and $\vpsi(r) = \vpsi_0$ (i.e.\ the Einstein
frame), then $\Omega=1$ and so the equations of motion
(\ref{eqn:parteoms3}) reduce to the familiar forms for timelike
geodesics in the equatorial plane $\theta=\pi/2$ of the line-element
(\ref{eqn:genform}) \cite{GRbook}. In the special case of the SdS
metric, where (\ref{eqn:SdSsoln}) holds, one therefore recovers the rotation
curve (\ref{eqn:SdSrotcurve}), which has no flat region.

Suppose one now performs a conformal transformation
$\tilde{g}_{\mu\nu}(x) = \Omega^2(r) g_{\mu\nu}(x)$ of the metric
(\ref{eqn:genform}) and also brings the angular part back into the
standard form in (\ref{eqn:genform}) by making the (radial) coordinate
transformation $r' = r\Omega(r)$ to obtain $\tilde{g}'_{\mu\nu}(x') =
{X^\rho}_\mu{X^\sigma}_\nu \tilde{g}_{\rho\sigma}(x(x'))$, where
${X^\rho}_\mu = \partial x^{\prime\rho}/\partial x^\mu$. As discussed
in \cite{Hobson:cgrotpaper}, in so doing, one finds that the resulting
line-element again has the form (\ref{eqn:genform}), but expressed in
terms of the new radial coordinate $r'$ and the metric functions
\be
\tilde{A}'(r') = \Omega^2(r(r')) A(r(r')), \quad
\tilde{B}'(r') = f^2(r(r')) B(r(r')),
\label{eqn:newab}
\ee
where we have defined the function $f(r) \equiv
1+r\frac{d\ln\Omega(r)}{dr}$.  In this new conformal frame, the
massive particle equations of motion are again given by
(\ref{eqn:parteoms3}), but with the replacements $r \to r'$, $A(r) \to
\tilde{A}'(r')$ and $B(r) \to \tilde{B}'(r')$. On substituting the
expressions $r'=r\Omega(r)$ and (\ref{eqn:newab}) into these equations
of motion, however, one finds after a short calculation that one
obtains precisely the {\it original} equations of motion
(\ref{eqn:parteoms3}) with $\Omega=1$, thereby recovering the particle
dynamics in the Einstein frame. Thus, for example, if $r = r(\phi)$ is
the orbit equation for a particle in the equatorial plane
$\theta=\pi/2$ in the Einstein frame, then the orbit equation in the
new conformal frame is given simply by $r' = r'(r(\phi))$.

This finding therefore eliminates, as it must, any ambiguity whereby
physical predictions appear to depend on the conformal frame in which
the calculation is performed. Specialising to the case where
(\ref{eqn:SdSsoln}) holds, this further demonstrates as unwarranted
the recent claims in the literature \cite{Li2019,Modesto2021} that one
may obtain flat galaxy rotation curves by conformally-rescaling the
SdS metric. It is worth pointing out that these claims arise from the
use instead of the equations of motion
(\ref{eqn:parteoms1}--\ref{eqn:parteoms2}), which are expressed in
terms of the parameter $\xi$, but where the latter is interpreted as
the particle proper time and implicitly assumed to be invariant under
conformal transformations, despite having a Weyl weight $w(\xi)=1$.
In that case, on following an analogous procedure to that we have
described above, one arrives at the erroneous conclusion that one does
not recover the particle dynamics in the Einstein frame and, more
generally, that particle trajectories depend on the conformal frame in
which they are calculated, which contradicts conformal invariance.

Although we have demonstrated that, when using the equations of motion
(\ref{eqn:parteoms3}) expressed in terms of the appropriate
conformally-invariant proper time $\tau$, one cannot obtain flat
galaxy rotation curves by {\it any} conformal rescaling of the SdS
metric, it is worth discussing briefly the particular rescaling
considered in \cite{Li2019,Modesto2021}. It is suggested in
\cite{Modesto2021} that there are physical reasons to require 
a metric of the form (\ref{eqn:genform}) to satisfy
the special condition $A(r)= B(r)$,
%that for a metric of the form (\ref{eqn:genform})
%to (i) preserve the null-energy condition for an ideal fluid matter
%source; (ii) ensure that the speed of light is constant in the empty
%space surrounding a point mass; and (iii) preserve the homogeneity of
%the universe on large scales, 
%one requires the special condition $A(r)
%= B(r)$ to hold
of which the SdS metric (\ref{eqn:SdSsoln}) is an example.\footnote{We
  discuss the wider implications of the gauge choice $A(r)=B(r)$ in
  \cite{ANL-MPH}, and in particular describe how, in fact, it is not
  only unnecessary, but also distorts the scaling properties of
  variables, thereby making it extremely difficult to identify
  `intrinsic' $\vpsi$-independent quantities that may be used for
  performing all calculations, including the derivation of the
  geodesic equations.} As shown in \cite{Hobson:cgrotpaper}, the
relations (\ref{eqn:newab}) imply that in order to preserve this
special condition, such that $\tilde{A}'(r') = \tilde{B}'(r')$, one
requires the conformal rescaling to have the unique form $\Omega(r) =
(1-ar)^{-1}$, where $a$ is an arbitrary constant, in which case $r' =
r/(1-ar)$ (or, equivalently, $r = r'/(1+ar')$ and $\Omega'(r') \equiv
\Omega(r(r')) = 1+ar'$). This matches the conformal rescaling and
coordinate transformation adopted in \cite{Li2019,Modesto2021} for
$a>0$. As shown in
\cite{Brihaye2009,Sultana2017,Horne2016,Hobson:cgrotpaper}, however,
these transformations convert the SdS metric into the
Mannheim--Kazanas metric \cite{Riegert1984,Mannheim1989}, so that the
claim in \cite{Li2019,Modesto2021} that one obtains flat galaxy
rotation curves in this conformal frame is merely a restatement of the
long-standing claims that conformal gravity predicts such rotation
curves
\cite{Mannheim1993b,Mannheim1997,Mannheim2011,Mannheim2012,Mannheim2017,OBrien2018},
although both claims are unjustified, as we have shown above.  It is
also worth noting that $r' \to \infty$ as $r \to 1/a$, so that the
$r'$ coordinate patch covers only a finite subset of the original $r$
coordinate patch. Indeed, it is straightforward to show that only a
finite interval of proper time $\tau$ is required for particle to
travel radially from some radius $r'=r'_0 > 2GM$ to $r' =
\infty$. This contradicts the claim in \cite{Modesto2021} that the
$r'$ coordinate patch is geodesically complete, which is based on the
fact that to reach $r' = \infty$ requires an infinite interval of the
parameter $\xi$, which is again mistakenly interpreted as the particle
proper time.

So far, our analysis has been limited to static, spherically-symmetric
systems, but our finding above that the particle dynamics is
independent of the choice of conformal frame is, in fact, entirely
general, as one might expect. As shown in \cite{Hobson:cgrotpaper}, the
action (\ref{ppaction3}) leads to massive particle equations of
motion in any conformal frame that are given by
\be
u^\sigma {u^{\mu}}_{;\sigma}  =  (g^{\mu\sigma} - u^\mu
u^\sigma)\vpsi^{-1}\,\partial_\sigma\vpsi,
\label{eqn:utrans3}
\ee
where the semi-colon denotes the standard Riemannian spacetime
covariant derivative ${u^\mu}_{;\sigma} = \partial_\sigma u^\mu +
\chssk{\mu}{\rho\sigma}u^\rho$.  Since the action (\ref{ppaction3})
is conformally invariant, these equations of motion are
covariant under conformal transformations, but are
not manifestly so.  If one uses local scale invariance to impose the
Einstein gauge $\vpsi = \vpsi_0$ (where, if desired, one can set
$\vpsi_0$ to unity without loss of generality), then
(\ref{eqn:utrans3}) reduces to
\be
u^\sigma {u^{\mu}}_{;\sigma}  \doteq  0,
\label{eqn:utrans4}
\ee
where $\doteq$ denotes that the equality holds only in a specific
gauge. Thus, in the Einstein gauge, a particle moving only under
gravity follows a geodesic of the metric $g^{\rm E}_{\mu\nu}$ in this frame,
as we already noted above for the special case of a static,
spherically-symmetric system.  As described in \cite{Lasenby:eWGTpaper},
however, it is unnecessary to break the scale symmetry by adopting a
particular gauge, since one may instead work in terms of
scale-invariant variables. Suppose in some arbitrary gauge, the metric
and scalar field are related to those in the Einstein gauge by
$g_{\mu\nu} = \Omega^2 g^{\rm E}_{\mu\nu}$ and $\vpsi =
\Omega^{-1}\vpsi_0$.  As mentioned above, one should identify $d\tau =
(\vpsi/\vpsi_0)\,d\xi = \Omega^{-1}\,d\xi$ as the interval of particle
proper time along its worldline. This leads one to define the
scale-invariant 4-velocity
\be
\hat{u}^\mu \equiv \nd{x^\mu}{\tau} = \nd{\xi}{\tau}\nd{x^\mu}{\xi}
=\left(\frac{\vpsi}{\vpsi_0}\right)^{-1}u^\mu = \Omega u^\mu,
\ee
which clearly has Weyl weight $w=0$. One may also define the
scale-invariant metric $\hat{g}_{\mu\nu} \equiv
(\vpsi/\vpsi_0)^2g_{\mu\nu} = \Omega^{-2}g_{\mu\nu}$
and its associated Christoffel connection
\bea
\widehat{\textstyle\chssk{\mu}{\rho\sigma}} & = & \tfrac{1}{2}\hat{g}^{\mu\nu}(\partial_\rho
\hat{g}_{\nu\sigma} + \partial_\sigma \hat{g}_{\rho\nu} - \partial_\nu
\hat{g}_{\rho\sigma}),\nonumber \\
& = & \textstyle\chssk{\mu}{\rho\sigma} + \vpsi^{-1}
(\delta^\mu_\rho \partial_\sigma \vpsi + \delta^\mu_\sigma\partial_\rho
\vpsi - g_{\rho\sigma}g^{\mu\nu}\partial_\nu \vpsi).\phantom{AA}
\eea
It is then straightforward to show that (\ref{eqn:utrans3}) may be
written in terms of the above scale-invariant variables as
\be
\hat{u}^\sigma {\hat{u}^{\mu}}_{\phantom{\mu}\,\widehat{;}\sigma} = 0,
\label{eqn:utrans5}
\ee
where we have defined
${\hat{u}^\mu}_{\phantom{\mu}\,\widehat{;}\sigma} \equiv
\partial_\sigma \hat{u}^\mu +
\widehat{\chssk{\mu}{\rho\sigma}}\hat{u}^\rho$. Thus, irrespective of
the gauge and without breaking the scale symmetry, the scale-invariant
4-velocity $\hat{u}^\mu$, which is appropriately defined in terms of
the particle proper time, satisfies the geodesic equation of the
scale-invariant metric $\hat{g}_{\mu\nu} = \Omega^{-2}g_{\mu\nu} =
g^{\rm E}_{\mu\nu}$, which is equal merely to the metric in the
Einstein gauge. Thus, we arrive at the conclusion that, quite
generally, the particle dynamics is independent of the choice of
conformal frame, as expected, and moreover satisfies the weak
equivalence principle.

Finally, we conclude by noting that the above conclusion applies not
only to conformally-invariant gravity theories defined on Riemannian
spacetimes, but also to those defined on more general Weyl--Cartan
spacetimes, which include Weyl, Riemann--Cartan and Riemannian
spacetimes as special cases. In a Weyl--Cartan ($Y_4$) spacetime, the
covariant derivative $\nabla_\mu = \partial_\mu +
{\Gamma^\sigma}_{\rho\mu} {\matri{X}^\rho}_\sigma $, where
${\Gamma^\sigma}_{\rho\mu}$ is an affine connection and
${\matri{X}^\rho}_\sigma$ are the $\mbox{GL}(4,R)$ generator matrices
appropriate to the tensor character of the quantity to which $\nabla_\mu$ is
applied. In particular, in a $Y_4$ spacetime, $\nabla_\mu$ satisfies
the semi-metricity condition
\be
\nabla_\sigma g_{\mu\nu} = -2B_\sigma g_{\mu\nu},
\label{wgtsemimet1}
\ee
where $B_\mu$ is the Weyl potential (and we have included a factor of
$-2$ for later convenience). On performing the simultaneous conformal
(gauge) transformations $g_{\mu\nu} \to \Omega^2(x)g_{\mu\nu}$, $B_\mu
\to B_\mu - \partial_\mu\ln\Omega(x)$, the condition
(\ref{wgtsemimet1}) is preserved \cite{Weyl1918}. From
(\ref{wgtsemimet1}), the connection is given by
\begin{equation}
{\Gamma^\lambda}_{\mu\nu} = {\textstyle\chssk{\lambda}{\mu\nu}} + \delta^\lambda_\nu B_\mu +
\delta^\lambda_\mu B_\nu - g_{\mu\nu}B^\lambda + {K^{\ast\lambda}}_{\mu\nu},
\label{weylaffinec}
\end{equation}
where ${K^{\ast\lambda}}_{\mu\nu}$ is the $Y_4$ contortion tensor,
which is given in terms of (minus) the $Y_4$ torsion
${T^{\ast\lambda}}_{\mu\nu} = 2{\Gamma^\lambda}_{[\nu\mu]}$ by
${K^{\ast\lambda}}_{\mu\nu}=-\tfrac{1}{2}({T^{\ast\lambda}}_{\mu\nu}-
{{{T^\ast}_\nu}^{\lambda}}_\mu + {{T^\ast}_{\mu\nu}}^{\lambda})$ (the
asterisks and the sign of the torsion are consistent with the usual
notation adopted in Weyl gauge theory
\cite{Blagojevic2002,Lasenby:eWGTpaper}). The matter action adopted in
$Y_4$ spacetime typically has the same form as that in
(\ref{eqn:sipgt0lm}) {\it et seq.}, but with the replacements
$\chssk{\lambda}{\mu\nu} \to {\Gamma^\lambda}_{\mu\nu}$ and $\partial_\mu
\to \partial_\mu^\ast = \partial_\mu + wB_\mu$, where $w$ is the Weyl
weight of the field being differentiated. As shown in
\cite{Hobson:scepaper2}, however, the corresponding action for a
spin-$\tfrac{1}{2}$ point particle is again equivalent to
(\ref{ppaction3}), which thus yields the equations of motion
(\ref{eqn:utrans3}). Equivalently, as shown in
\cite{Lasenby:eWGTpaper}, the equations of motion derived from the
point particle action in $Y_4$ spacetime may be rewritten directly as
(\ref{eqn:utrans3}). In either case, one thus arrives at the same
conclusions as reached above for Riemannian spacetimes.

%%%%%%%%%%%%%%%%%%%%%%%%%%%%%%%%%%%%%%%%%%%%%%%%%%%%%%%%%%%%%%%%%%%%%%%%%%%%%
%\bigskip
%\begin{acknowledgments}
%We thank Philip Mannheim for his comments on the original version of
%this paper, and the anonymous referee for several useful suggestions.
%\end{acknowledgments}
%%%%%%%%%%%%%%%%%%%%%%%%%%%%%%%%%%%%%%%%%%%%%%%%%%%%%%%%%%%%%%%%%%%%%%%%%%

\end{document}